# Chemical nanomachining of silicon by gold-catalyzed oxidation

*J.T. Robinson[1,2], P.G. Evans[3], J.A. Liddle[2], O.D. Dubon[1,2]*

[1]Department of Materials Science and Engineering, University of California, Berkeley, CA 94720.

[2]Lawrence Berkeley National Laboratory, Berkeley, CA 94720.

[3]Department of Materials Science and Engineering, University of Wisconsin, Madison, WI 53706.

* Corresponding Author E-mail: oddubon@berkeley.edu

**ABSTRACT**

A chemical nanomachining process for the rapid, scalable production of nanostructure assemblies from silicon-on-insulator is demonstrated. The process is based on the spontaneous, local oxidation of Si induced by Au, which is selectively evaporated onto the Si surface. The Au-catalyzed oxide forms a pattern that serves as a robust mask for the underlying Si, enabling the use of simple wet chemistry to sculpt arrays of nanostructures of diverse shapes including rings, pillars, wires, and nanopores. The remarkable simplicity of this chemical nanomachining process makes it widely accessible as an enabling technique for applications from photonics to biotechnology.

The fabrication of semiconductor nanostructures and their organization into functional macroassemblies and metamaterials remain a fundamental challenge in nanoscience and nanotechnology, requiring new strategies in materials processing. Successful approaches are likely to combine "top-down" patterning technologies that are very precise and highly reproducible with spontaneous self-organizing "bottom-up" processes that minimize the need to directly sculpt and manipulate each nanoscale element, one at a time[1-4]. Here we demonstrate a route for the fabrication of silicon nanostructure assemblies in which top-down and bottom-up approaches naturally converge. The process is based on the oxidation of Si catalyzed by Au. The application of a Au pattern to a Si surface provides spatial control over the oxidation process, which occurs exclusively at and in the immediate vicinity of surface sites where Au is present. The spontaneously formed, self-patterned oxide is sufficiently robust to serve as a mask for the production of nanostructures by simple wet etching. We describe how this chemical nanomachining process can be used to produce a variety of nanostructure arrays from silicon-on-insulator (SOI).

In the presence of Au, Si can oxidize by two simultaneous processes. A silicon substrate that is covered by a uniform Au film oxidizes even at room temperature by the diffusion of Si through the Au film in an oxidizing atmosphere including air [5-8]. The formation of an alloyed Si-Au interlayer enhances oxidation by the Au-induced modification of the hybridization state of Si atoms that have diffused toward the surface. This interlayer, however, limits the extended growth of the oxide by inhibiting further diffusion of Si from the underlying substrate. Thus, the resulting oxide may be removed by rinsing in hydrofluoric acid, but subsequent exposure of the HF-etched surface to air does not restore the Si oxide[6].

A second oxidation process arises from the potential difference between Au and Si, which drives an electrochemical reaction. If a Au pattern is deposited onto a clean Si surface, Si in the immediate vicinity of the Au-covered regions electrochemically oxidizes, or anodizes, upon exposure to air (more specifically water vapor in the air). Removing the Si native oxide by HF rinsing prior to the application of the Au pattern is an important initial step in this process as it allows for a conducting path for electrons between Si and Au. The anodic oxide forms along the perimeter of the patterned Au features, and unlike the oxidation of Si through a Au film, rinsing in HF and subsequent re-exposure to an oxidizing atmosphere result in the restoration of the anodic oxide. This intriguing difference *between the formation* of the anodic oxide and the Si oxide formed by Si diffusion through the Au—henceforth referred as the "through-Au oxide"—allows for machining of SOI into unique nanostructure assemblies. We note that Si can be anodized locally in air without Au by applying an external potential, for example, with a scanning probe tip[9,10].

The fabrication of Si nanostructures by chemical nanomachining begins by rinsing (001)-oriented, 105 nm-thick SOI in a solution of $HF:H_2O$ (1:1), which hydrogen terminates the SOI surface. A Au pattern is applied by evaporating Au through the windows of a silicon nitride stencil mask that is placed in direct contact with the SOI. Spontaneous oxidation at and around each Au feature upon exposure to air yields a Si surface with a protective oxide pattern. Etching the SOI with aqueous solutions of HF and/or KOH yields Si nanostructures whose geometries depend exclusively on the shapes of the Au features (determined by the window geometries of the stencil mask) and sequence of etching solutions used. The process is depicted in Figure 1a for the formation of pillars and rings.

The atomic force microscopy (AFM) height images of pillars and rings presented in Figs. 1b and 1c, respectively, were taken from regions that contain tens of thousands of nearly identical structures. Pillars were formed from SOI that is decorated with an array of Au squares by etching in a KOH solution, which does not attack $SiO_2$. Rings rather than pillars were produced simply by immersing a similarly Au-patterned SOI sample in HF prior to etching with KOH. Extensive arrays are readily detected by optical diffraction as shown in Fig. 1d for the case of ring arrays. Interestingly, the inner perimeters of the rings retain the initial shape of the Au feature as shown in Fig. 1e.

The etching processes leading to the formation of pillars and rings are elucidated by analyzing the effect of wet etching on a patterned Au square and the surrounding Si surface. A series of AFM height images and line profiles illustrating the response of Au-patterned Si(001) to various etching sequences is presented in Figure 2. The as-deposited Au square appearing in Fig. 2a has a well-defined square footprint but is taller than expected from the nominal thickness of 1 nm of the deposited Au. This enhancement in height is attributed to the diffusion and subsequent oxidation of Si within the Au square[6]. The effect of immersing Au-patterned Si(001) in KOH for two minutes is shown in Figure 2b. Silicon is etched in the unprotected regions; however, both the anodic oxide corona that surrounds the Au square and the through-Au oxide are sufficiently robust to mask the underlying Si from KOH etching. The height profile reveals the persistence of the Au square after etching. Similarly, the tops of pillars formed from SOI are each capped with a Au square and the oxide that forms within and around it.

The result from rinsing Au-patterned Si in HF is illustrated in Fig. 2c. The depression precisely at the Au square reflects the removal of the through-Au oxide by the HF solution. Etching extends slightly beyond the perimeter of the Au square indicating that the anodic oxide is etched by HF as well. Brief exposure to HF followed by KOH etching of Au-patterned Si yields the topography shown in Fig. 2d. In this case, HF removes the through-Au oxide, and subsequent rinsing in KOH etches Si that is not only beyond the outer perimeter of the anodic oxide corona but also directly where Au was deposited. This indicates that the anodic oxide is restored even after HF etching while the through-Au oxide does not regrow due to the presence of a Si-Au interlayer[6]. We have found no difference in the nature of the through-Au and anodic oxides; rather, the major difference between the two arises in the manner by which they form.

The Au-enhanced oxidation of Si at the nanoscale has been reported in the context of the vapor-liquid-solid (VLS) process for Si nanowire synthesis[11]. It was found that in the presence of oxygen the Au catalyst droplet located at the growing tip of a Si nanowire enhances the oxidation of the nanowire surface. These results suggest that Au nanoparticles may be used to form even more refined oxide nanopatterns on Si.

When the Au-squares are sufficiently close together, the anodic oxide coronas around the Au squares overlap to form a continuous surface oxide. In this case, etching Au-patterned SOI with HF followed by KOH produces a continuous Si film with holes rather than the rings observed for the case of more widely spaced Au squares. Figure 3a shows an AFM phase image of the border region between arrays of chemically nanomachined holes of different size and pitch formed in the silicon device layer of the SOI sample. The size and arrangement of the holes are programmed through stencil mask design, providing a simple route for the production of novel two-dimensional photonic structures[12].

The height profile along the line indicated in Figure 3b is shown in Figure 3c. The hole was produced from a 75×75 nm$^2$ Au square and has a dimension smaller than 20 nm across the bottom opening. Additional refinement of nanopore formation can be expected through a combination of optimizing the dimensions of the Au squares, the parameters of the etching in KOH solution, and the thickness of the Si layer. The further reduction of hole size to fewer than 10 nm opens opportunities for the fabrication of arrays of nanopores that may be used as filters or for biomolecule analysis[13,14].

The shape of each gold feature provides a precise tool to sculpt Si into a variety of structures through this chemical nanomachining process. For example, connected Si nanowires were obtained by etching SOI patterned with Au lines 200 nm in width in HF and KOH. Having the appearance of a stretched rubber band, each connected wire pair is 25 μm long, a length limited only by the long dimension of the stencil mask window. Segments of these structures are presented in Figs. 4a and 4b. The double wires have the same height as the original 105 nm-thickness of the Si device layer (Fig. 4c); the Si has been etched completely to the oxide in both the surrounding and central regions of these structures. Similarly to the rings and pillars, etching a Au line with KOH alone leaves the Au-capped Si intact and leads to a single wire with a width corresponding to that of the double wire structure.

Combinations of nanostructures of widely varying geometries can be formed. Figure 4d presents a section from an array of double wires and rings. Further functionality can be gained by releasing the nanostructures from the original SOI substrate in a subsequent HF rinse and transferring them onto other substrates. The achievement of enhanced device functionalities through the integration of Si-based elements onto diverse substrates has been recently demonstrated[15,16].

The appeal of directed local oxidation for nanostructure sculpting is not limited to the Si-Au system alone. Platinum, for example, catalyzes the local anodization of Si. Other semiconductors such as GaAs also display metal-enhanced oxidation, creating opportunities for the chemical nanomachining of nanostructures from light emitting materials[17]. The availability of other patterning methods such as soft lithography creates further possibilities for the production of metal patterns, which could be used to produce more complex, functional nanoassemblies.

Acknowledgments. The authors thank T.M. Devine and C. Kumai for insightful discussions and O.G Schmidt for use of facilities at the Max-Planck-Institute Stuttgart. O.D.D. acknowledges support from the National Science Foundation under contract number DMR-0349257. J.T.R. acknowledges support from the IBM PhD fellowship program. The work at the Lawrence Berkeley National Laboratory was supported in part by the Director, Office of Science, Office of Basic Energy Sciences, Division of Materials Sciences and Engineering, of the U.S. Department of Energy under Contract No. DE-AC02-05CH11231. P.G.E. acknowledges support by the National Science

Foundation through the University of Wisconsin Materials Research Science and Engineering Center, grant number DMR-0520527.

**Figure Captions**

Figure 1 (a) Schematic of chemical nanomachining of Si pillars and rings. (b,c) AFM images of pillars and rings formed on 105 nm SOI decorated with an array of nominally 1 nm-thick Au squares, each of which was 200×200 nm$^2$. The buried oxide exposed after etching is indicated by the purple coloring. (d) Optical diffraction from arrays of differently spaced nanorings. Each region measures 125×125 μm$^2$ and contains 1 to 2×10$^4$ rings. (e) AFM images of the inner perimeters of the ring structures formed from an initial Au squares and triangles.

Figure 2  AFM images (scale bars=200 nm) of (a) an as-deposited Au square and the surrounding area of the Si surface, (b) Au square after etching in KOH, (c) the Au-square presented in (a) after etching in HF; and (d) a Au-square after etching in HF followed by KOH.  To the right of each height image is a corresponding linear height profile taken along the lines indicated.  The width of the yellow band indicates the approximate lateral extent of the deposited Au square.  The height in the profiles is referenced approximately to the original silicon surface; therefore, etching appears as a negative height.

Figure 3 (a) AFM height image of chemically nanomachined holes in 105 nm-thick SOI. The image shows a section of the boundary between two arrays of holes of different size and spacing.  (b) AFM height image of a single hole machined from a Au square 75 nm in side length deposited on the SOI. (c) Height profile taken along the line in (b).  The sub-20-nm opening at the bottom of the hole is 105 nm from the surface indicating that the oxide layer has been reached. Extrapolation of the tapered sidewalls to the oxide layer is consistent with the sub-20 nm dimension of the hole.

Figure 4 (a) Scanning electron micrograph showing the ends of a Si double nanowire array, each double wire measuring 25μm in length. (b) Perspective AFM close-up and corresponding height profile, (c) taken along the line indicated. (d) AFM image showing Si ring and line structures formed on a Si substrate by etching in HF and KOH.

**References**

1. Salaita K. S.; Lee, S.W.; Ginger,D. S.; Mirkin, C. A. *Nano Lett.* **2006,** 6, 2493 .

2. Menke, E.J.; Thompson, M.A.; Xiang, C.; Yang, L.C.; Penner, R.M. *Nature Materials* **2006**, 5, 914.

3. Ji, R.; Lee, W.; Scholz, R.; Gösele, U.; Nielsch, K. *Adv. Mater.* **2006**, 18, 2593.

4. Kamins, T. I.; Williams, R. S. *Appl. Phys. Lett.* **1997**, 71, 1201.

5. Hiraki, A.; Nicolet, M-A; Mayer, J.W. *Appl. Phys. Lett.* **1971**, 18, 178.

6. Hiraki, A.; Lugujjo, E.; Mayer, J.W. *J. Appl. Phys.* **1972**, 43, 3643.

7. Cros, A.; Derrien, J.; Salvan, F. *Surf. Science* **1981**, 110, 471.

8. Derrien, J. *et al. Appl. Phys. Lett.* **1981**, 39, 915.

9. Dagata, J.A. *et al. Appl. Phys. Lett.* **1990** 56, 2001.

10. Snow, E.S. *et al. Appl. Phys. Lett.* **1994** 64, 1932.

11. Kodambaka, S.; Hannon, J. B.; Tromp, R. M.; Ross, F. M. *Nano Lett.* **2006**, 6, 1292.

12. Akahane, Y.; Asano, T.; Song, B.-S.; Noda, S. *Nature* **2003**, 425, 944.

13. Li, J. *et al. Nature* **2001**, 412,166.

14. Storm, A.J.; Chen, J.H.; Ling, X.S.; Zandbergen, H.W.; Dekker, C. *Nature Materials* **2003**, 2, 537.

15. Khang, D.-Y.; Jiang, H.; Huang, Y.; Rogers, J.A. *Science* **2006**, 311, 208.

16. Yuan, H.-C.; Ma, Z.; Roberts, M.M.; Savage, D.E.; Lagally, M.G. *J. Appl. Phys.* **2006**, 100, 013708.

17. Kubota, T.; Nakato, Y.; Yoneda, K.; Kobayashi, H. *Phys. Rev. B* **1997**, 56, 7428.

Robinson *et al.* Figure 1

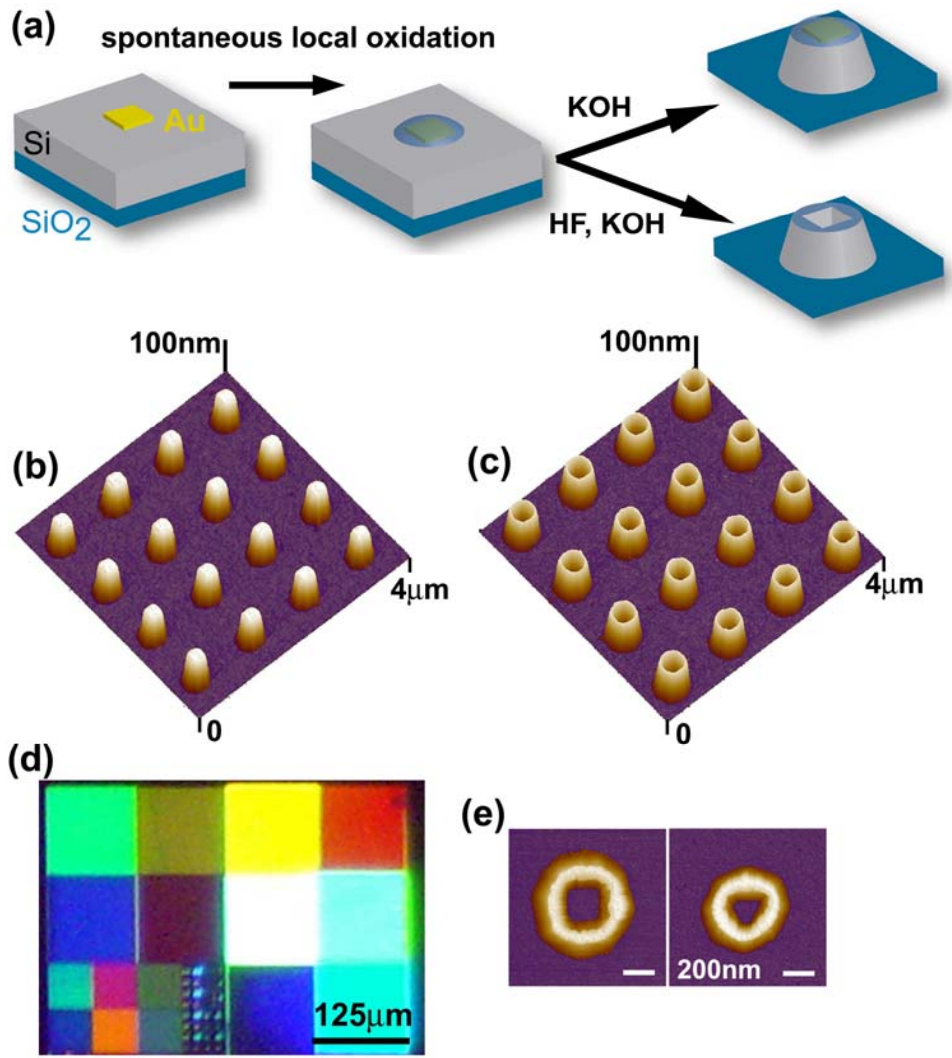

Robinson *et al.* Figure 2

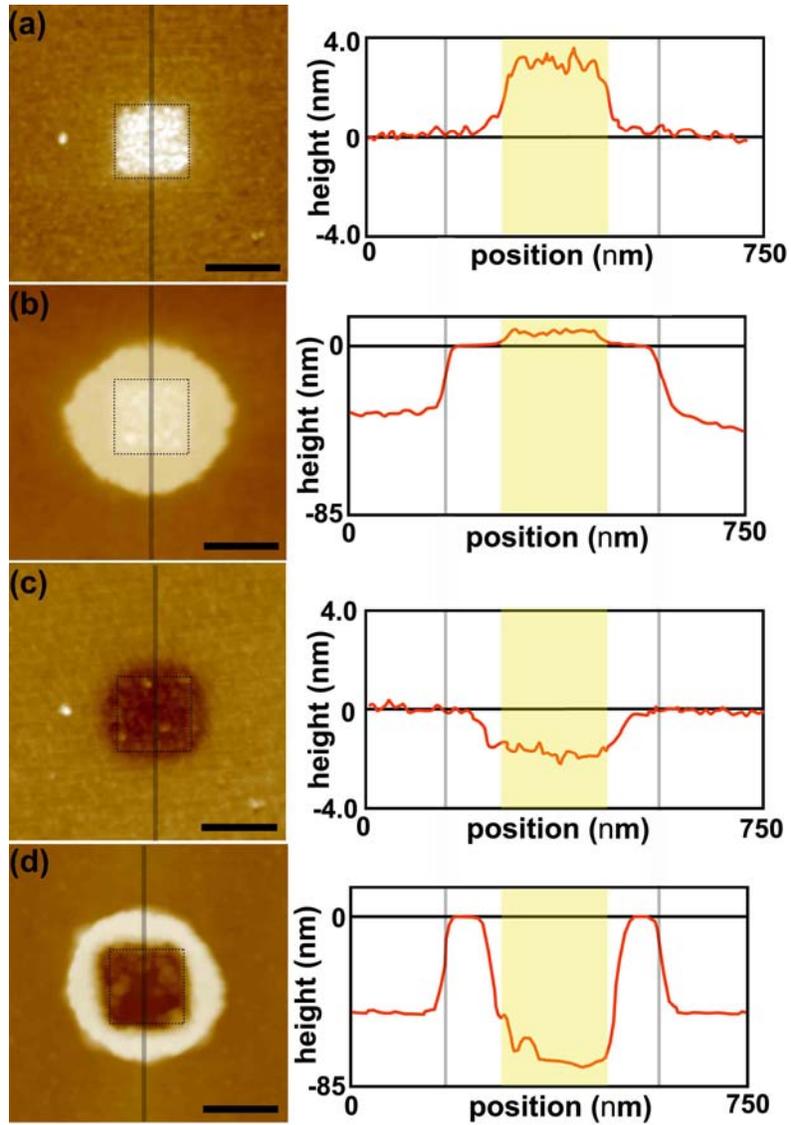

**Robinson *et al.* Figure 3**

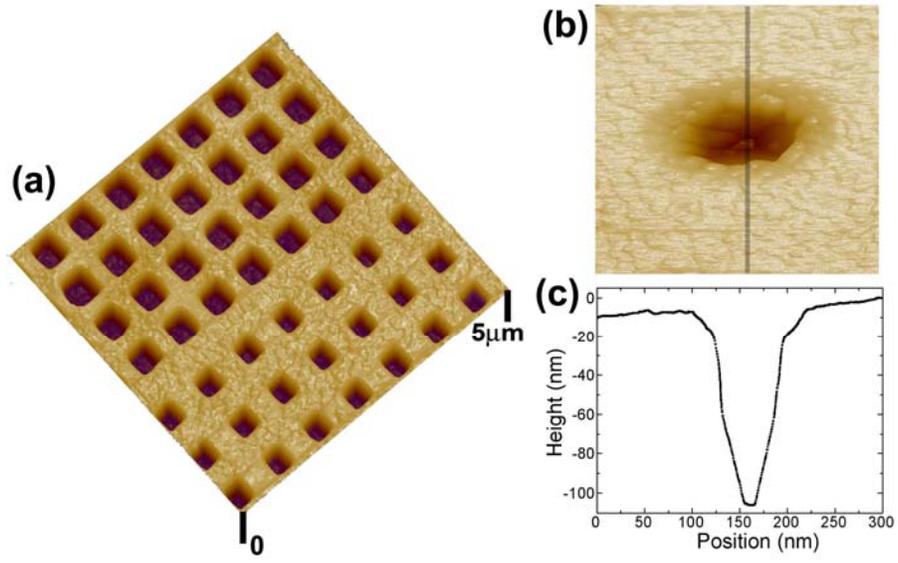

**Robinson *et al.* Figure 4**

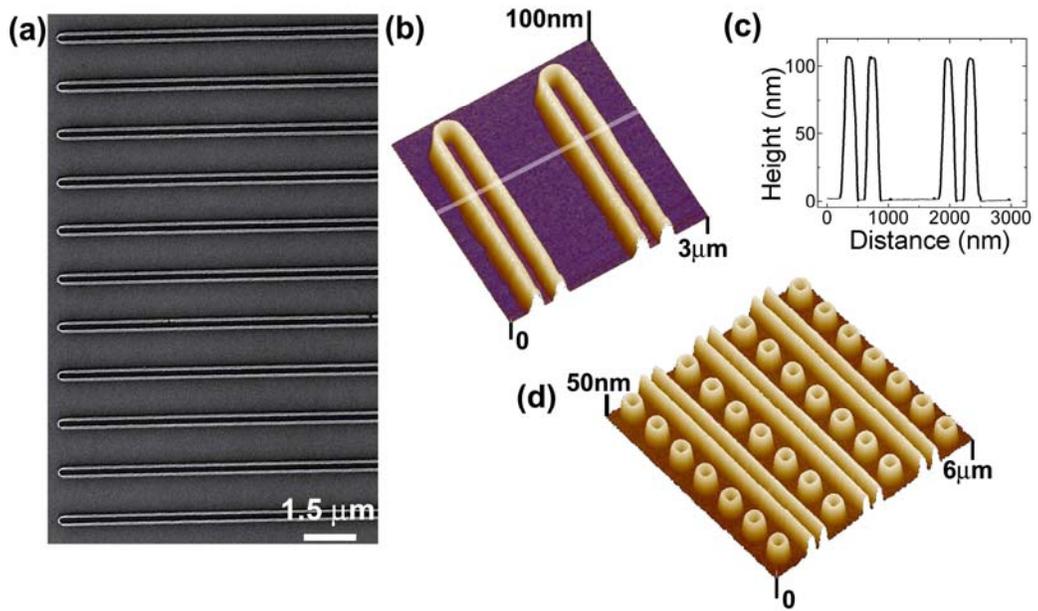